\documentclass[preprint,authoryear,12pt]{elsarticle}

\usepackage{epsfig}

\usepackage{amssymb}
\usepackage[%ps2pdf,%
a4paper=true,%
breaklinks=true,%
colorlinks=true,%
pdfauthor={First Author et al.},%
pdftitle={Template for manuscripts in Advances in Space Research}%
]{hyperref}
\def\mnras{Mon. Not. Royal Astr. Soc.\/}
\def\apj{Astroph. J.\/}
\def\apjl{Astroph. J. Lett.\/}

\def\apjs{Astroph. J. Suppl.\/}
\def\aap{Astron. Astroph. \/}

\def\nat{Nature}
\journal{Advances in Space Research}

\begin{document}

%%%%%%%%%%%%%%%%%%%%%%%%%%%%%%%%%%%%%%%%%%%%%%%%%%%%%%%%%%%%%%%%%%%%%%%%%%%%%
%% Frontmatter
\begin{frontmatter}


% Use the tnoteref command within \title and fnref within \author or \address for footnotes;
% use the corref command within \author for corresponding author footnotes;
% \author{Name\corref{cor1}\fnref{label2}}
% \address{Address\fnref{label3}}

\title{Techniques for Profile Binning and Analysis of Eigenvector Composite Spectra: 
Comparing H$\beta$\ and Mg{\sc ii}$\lambda$2800  as Virial Estimators \tnoteref{footnote1}}
\tnotetext[footnote1]{Invited talk presented on May 12, 2013 at the 9th SCSLSA held in Banja Kovilia\v{c}a, Serbia.}

% Use optional labels to link authors explicitly to addresses:

\author{Jack  W. Sulentic\corref{cor}}
\address{Instituto de Astrof{\'\i}sica de Andaluc{\'\i}a (CSIC), Granada, 18008, Espa\~na}
%\fntext[footnote3]{Additional information about the second and third authors}
\cortext[cor]{Corresponding author}
\ead{sulentic@iaa.es}

\author{Paola Marziani}
\address{INAF, Osservatorio Astronomico di Padova, Padova,35122, Italia}
%\fntext[footnote2]{Additional information regarding the corresponding author}
\ead{paola.marziani@oapd.inaf.it}

% Url can be given like this:
% \ead[url]{http://www.elsevier.com/wps/find/authorsview.authors/latex}

\author{Ascensi\'{on} del Olmo and Ilse Plauchu-Frayn\fnref{footnote2}}
\address{Instituto de Astrof{\'\i}sica de Andaluc{\'\i}a (CSIC), Granada, 18008, Espa\~na}

%\author{More Authors\fnref{footnote4}}
%\address{Address of the co-authors}
\fntext[footnote2]{Present address: IA-UNAM, Ensenada Campus, Ensenada, M\'{e}xico}
%\ead{more@email.addresses}

\begin{abstract}
We review the basic techniques for extracting information about quasar structure and kinematics  from the broad emission lines in quasars. We consider which lines can most effectively serve as virial estimators of black hole mass. At low redshift  the Balmer lines,particularly broad H$\beta$, are the lines of choice. For redshifts greater than 0.7 -- 0.8 one can follow H$\beta$ into the IR windows or find an H$\beta$ surrogate. We explain why UV C{\sc iv}$\lambda$1549 is not a safe virial estimator and how Mg{\sc ii}$\lambda$2800 serves as  the best  virial surrogate for H$\beta$\ up to the highest redshift quasar known at $z \approx 7$. We show how spectral binning in a parameter space context (4DE1) makes possible a more effective comparison of H$\beta$ and Mg{\sc ii}. It also helps to derive more accurate mass estimates from appropriately binned spectra and, finally, to map the dispersion in $M_\mathrm{BH}$\ and Eddington ratio across the quasar population. FWHM MgII is about 20$\%$ smaller than FWHM H$\beta$ in the majority of type 1 AGN requiring correction when comparing $M_\mathrm{BH}$\ estimates from these two lines. The 20$\%$ of sources showing narrowest FWHM H$\beta$ ($< 4000$ km s$^{-1}$) and strongest FeII ($R_\mathrm{Fe} \gtrsim 1.0$) emission (we call them bin A3-4 sources) do not show this FWHM difference and a blueshift detected in MgII for these sources suggests that FWHM H$\beta$ is the safer virial estimator for these extreme Eddington emitters.

\end{abstract}

\end{frontmatter}

\parindent=0.5 cm

%%%%%%%%%%%%%%%%%%%%%%%%%%%%%%%%%%%%%%%%%%%%%%%%%%%%%%%%%%%%%%%%%%%%%%%%%%%%%
%% Main text

\section{Introduction}

Quasars were discovered in 1963 as radio-loud blue-stellar sources showing very broad emission lines in their high redshifted spectra. Fifty years later we know that most quasars are radio-quiet and show a wide diversity of spectroscopic properties united under the umbrella of Active Galactic Nuclei (AGN). The large population of AGN showing broad emission lines can be viewed as a luminous high-accreting parent population extending from local Seyfert galaxies like NGC1068 to the quasar with highest known redshift at $z \approx$7.03. Spectroscopy is the fundamental method we employ to ``resolve'' the  central structure of line emitting gas (broad line region: BLR; see the review by \citealt{gaskell09}) surrounding the supermassive black hole thought to lie at the center of each AGN. We study the broad emission lines to infer BLR structure and kinematics. More recently we have begun using the width of select broad lines to estimate the mass of the black hole and which also gives us an estimate of the source Eddington ratio.

Broad emission lines characterise an important and dominant spectroscopic
class of high accreting active galactic nuclei (AGN). Most are accreting at a high
enough rate to fuel a reasonably stable broad line emitting region (BLR). The (likely) lowest
accreting sources among them e.g. NGC 1097 \citep{storchi-bergmannetal93} and 
NGC 4151  \citep{bonetal12} are less stable and show broad lines only part of the time. 
Some are obscured (by a torus?) and show broad lines only in polarized light 
e.g. NGC 1068 \citep{antonuccimiller85}. At low redshift, where a host galaxy can be seen, 
they are called Seyfert 1 galaxies and at higher redshift, where they are largely unresolved 
point sources, we call them Type 1 quasars. Seyferts 1's were long ago identified 
by Carl Seyfert \citep{seyfert43} while discovery of the quasars occured fifty years ago \citep{schmidt63}. Broad lines alone cannot define the AGN phenomenon because many subtypes lacking them are now known (e.g. BL Lacs, Type 2 AGN and LINERs). Our focus in this paper will
be exclusively on Type 1 AGN.

Whether hosted in a detectable galaxy or not, the region producing the
broad emission lines is spatially unresolved. Such source compactness, and associated
short variability timescales, are consistent with the hypothesis that the broad lines are
produced in a region no larger than a fraction of a light year in diameter and connected
with gas accreting onto a supermassive black hole \citep[e.g.][]{gaskellsparke86,padovanirafanelli88}. Spectroscopy therefore is, and will likely remain, the means by which we ``resolve'' this broad line region
(BLR). Broad line widths and shapes then provide the most direct clues about BLR
geometry and kinematics.All broad lines in an individual source do not necessarily show 
the same properties and any specific line can as well show striking diversity in different
quasars. The latter fact offers an immediate motivation for attempts to systematize 
measures of specific broad lines. 

In addition to providing clues about BLR structure and kinematics, broad lines are now 
used increasingly to estimate the mass of the 
central black hole which measure is of great importance both for models of BLR 
structure and (potentially) for cosmology (see contribution by Marziani \& Sulentic 
in this proceeding). Which lines can we use for black hole mass estimation and over 
what redshift range is a particular line useful? We seek lines that arise in a virialized 
medium which leads us to the Balmer lines at low redshift. H$\beta$\ is the most useful 
Balmer line for such studies because it usually shows only small shifts and 
asymmetries -- it thus appears to be a ``safe'' virial estimator for most Type 1 AGN. If it 
arises from a flattened distribution of emitting clouds or, even from a Keplerian accretion 
disk, as is often assumed, then the virial assumption is not unreasonable. H$\beta$\  suffers 
from contamination by nearby lines of Fe{\sc ii}, He{\sc ii}$\lambda$4686 and (narrow) 
[O{\sc iii}]$\lambda\lambda$4959,5007 but these can be reasonably modeled and 
subtracted to facilitate evaluation of line properties. H$\beta$\ is lost to optical 
spectroscopy in the redshift  range $z \sim 0.7 - 0.9$. It can be followed through the JHK 
infrared windows out to $z \approx 3.7$\ using suitably large telescopes although this approach 
is still rather costly in terms of telescope time. Estimates of black hole masses for large high redshift samples require an H$\beta$\ surrogate. Two lines, C{\sc iv}$\lambda$1549  and Mg{\sc ii}$\lambda$2800 \citep[e.g., ][]{mclurejarvis02}, are the best candidates in terms of line strength and low levels of contamination.

High ionization C{\sc iv}$\lambda$1549 is so dangerous that many prefer to avoid it entirely 
\citep{sulenticetal07,netzeretal07}. Yes ironically, various studies suggest that it is virialized \citep{gaskell88,petersonwandel99}. CIV often shows profile blueshifts and asymmetries interpreted as signatures of winds or outflows \citep{gaskell82,gaskellgoosmann13}. These C{\sc iv}$\lambda$1549 distortions are often seen in quasars where H$\beta$\ is most symmetric and well behaved. The problem with CIV is therefore not the issue of virialization but the likelyhood that the line emission arises highly anisotropically in a structure with unknown geometry. 
FWHM C{\sc iv}$\lambda$1549 does not show a clear correlation with FWHM H$\beta$ the low 
$z$\ virial estimator of choice and this  precludes a simple geometric relation between  C{\sc iv}$\lambda$1549  and H$\beta$ as well as Mg{\sc ii}$\lambda$2800 emissions \citep{shenetal08}. Thus C{\sc iv}$\lambda$1549 is falling from (actually it was never in) favor, and Mg{\sc ii}$\lambda$2800 is replacing it, as the preferred high redshift virial estimator. 
This paper compares H$\beta$ with Mg{\sc ii}$\lambda$2800 and presents evidence in support 
of the latter line as a virial estimator for 80-90\% of high redshift quasars--all except the highest accretors. This paper, and the emergence of Mg{\sc ii}$\lambda$2800, are largely due 
to the advent of the SDSS which provides Mg{\sc ii}$\lambda$2800 spectra for thousands 
of quasars. Most importantly for 500+ sources where both H$\beta$ and 
Mg{\sc ii}$\lambda$2800 can be measured in the same spectra. Low 
ionization Mg{\sc ii}$\lambda$2800 is reasonably symmetric and unshifted like H$\beta$ making 
it the best candidate as a surrogate. Mg{\sc ii}$\lambda$2800 has even been measured
(K band) in the quasar with highest current redshift ($z \approx$ 7.035; \citealt{mortlocketal11}).

\section{Estimating Black Hole Mass}

When estimating black hole mass one requires both a velocity dispersion and a radius of the BLR.
If the virial assumption is valid then $M_\mathrm{BH} = f (\delta v)^2r/G$. $f$\ is a parameter 
that accounts for line-of-sight effects on the line profile due to geometry and kinematics. It may impose the ultimate limitation on the accuracy of $M_\mathrm{BH}$ estimates. 
We do not know the correct value of $f$\ and, more importantly, how much it changes 
across the quasar population. The value of $r$\ (the effective BLR radius) can be directly estimated using reverberation techniques as it simply involves the delay time in the 
response of an emission line to continuum changes. Reverberation mapping is prohibitively expensive in terms of required telescope time meaning that only
about 60 of the brightest nearby (and hence predominantly low luminosity) sources have been
reverberation mapped. The reverberation radii tend to correlate with measures of
source luminosity leading to a radius -- luminosity relation \citep{dibai77,koratkargaskell91,kaspietal05} $r \propto L^{\alpha}$
($\alpha \approx $0.5) which can provide, via extrapolation, $r$ estimates for all sources 
where $L$\ can be reliably measured. Note that we have only a vague notion of what $ r$\ means 
since the BLR is unlikely to be a shell surrounding the central continuum source \citep[][and Alenka Negrete's contribution in this proceedings]{negreteetal13}.

Measures of $\delta v$ (the virial velocity dispersion) come from direct measurement of a line 
that is considered a reliable virial estimator. If a given broad
line is reasonably symmetric and unshifted we adopt FWHM H$\beta$\ (or 
FWHM Mg{\sc ii}$\lambda$2800?) as the virial estimator. We do this because a 
symmetric line without inflections implies a reasonably coherent single source 
of line emission involving motions centered around
the quasar rest frame (usually inferred from the redshift of narrow lines
like [O{\sc iii}]$\lambda\lambda$4959,5007). This is the simplest expectation for a bound 
virialized emitting region.  If the line profile shows inflections we do not know
what to do because inflections imply spectroscopic resolution of the BLR. The
spatially unresolved source could involve multiple kinematically distinct emitting components --a
binary black hole comes to mind -- if each black hole is  surrounded by a BLR. This is
perhaps less likely than such profile complexities indicating stratification of the
emitting region and/or radial motions of all or part of the line emitting gas (winds,
outflows or infall). Line profile inflections are a valuable clue telling
us that multiple emitting components are present and that all or part of the line
cannot be safely regarded as a virial estimator. The study of profile inflections is still in its infancy because they require high resolution (like SDSS) and high s/n ( like only a few hundred SDSS) spectra. 

Three broad line components have
been identified in H$\beta$\ \citep{marzianietal10} and presumably only one of them
is/might be a virial estimator. The situation is not so bad because
many sources show only a single component  ($\sim$50\% of quasars in the case of 
H$\beta$). Naturally for the other 50\%\ we adopt the least shifted broad component
for black hole mass estimation. The larger blueshifts and asymmetries observed in
C{\sc iv}$\lambda$1549 represent a more serious problem--in this case we are not 
even sure if any part of the line can be trusted. Another problem with 
C{\sc iv}$\lambda$1549 involves the difficulty of subtracting a narrow line component 
which if uncorrected will cause underestimation of the black hole mass 
\citep{sulenticmarziani99,sulenticetal07}. C{\sc iv}$\lambda$1549 is not the road
to more accurate black hole masses. Note that we do not want to use many 
different lines  for virial mass estimation because this simply adds uncertainty 
when we compare  them and try to tie them together over different redshift ranges. 
That is why it is likely that H$\beta$ and Mg{\sc ii}  represent the 
best and safest set of virial estimators that can cover the full redshift range.  In 
other words we do not need or want C{\sc iv}$\lambda$1549 especially as 
a (more uncertain) third estimator.

\section{FWHM H$\beta$ vs. $\sigma$\ as the Virial Estimator?}

The second moment of the line profile $\sigma$ (also called the line
dispersion) has sometimes been favored over  profile width as a more physical 
and/or reliable virial measure. The relative merits of the two measures as 
virial estimators were discussed in \citet{collinetal06}. Both FWHM
and $\sigma$ are simply numbers (albeit one in units of line-of-sight velocity)
and 1) if the line profile of the adopted virial
estimator were similar from source-to-source and 2) if  a uniform range of profile 
width was observed, then the two numbers should be fully equivalent (allowing only for 
the possibility that one might be measured more accurately  than the 
other, \citealt{petersonetal04}) Unfortunately broad lines sometimes show inflections 
indicating  2 or 3 emitting components. Since they cannot all be virial estimators neither a FWHM nor
$\sigma$ measure for the full profile should be used. The line profile can
at least be used to model the line components \citep{marzianietal10} while $\sigma$\ 
is of no use and has no meaning. If one models
the individual line components and isolates a FWHM measure for the
component most likely to be a virial
estimator then  there seems little additional value in computing $\sigma$\
 and using it instead of FWHM as the virial measure. Profile modeling is the road to
more accurate black hole mass estimates in the future but it can only be 
applied if spectral S/N is high enough. A S/N$\geq$20\
computed in the continuum near, but not including, the adopted virial
estimator is a useful rule of thumb.

In addition to broad line inflections there is often the well known 
inflection between broad  and narrow line components. This is especially important when 
using H$\beta$\ as the virial estimator.
When the inflection is clear--especially for sources where broad H$\beta$
shows FWHM $>$ 3000 km s$^{-1}$ -- it can
be used to guide subtraction of  the narrow component. Black hole mass estimates
from a broad line without correction for the narrow component will be 
underestimates. Again 
reasonable spectra S/N is desirable if one wishes
to obtain an estimate with less than 1dex uncertainty.

\section{Contextualization of Broad Line Properties: 4DE1}

Fortunately narrow line and broad line inflections do not occur randomly
but show trends. What we need is a context in which to interpret broad line spectra. 
The PCA analysis of high S/N PG spectra \citep{borosongreen92}
opened the door to a new era of source contextualization. These results
provided the first hints that all quasar spectra are {\em not} self similar 
and that indiscriminate averaging of quasar spectra \citep{vandenberketal01} -- no matter
how tempting -- is not  the path to progress. We have built upon the PG
results using larger samples of
quasars and have proposed a 4D Eigenvector 1 (4DE1) parameter space
\citep{sulenticetal00a} involving correlations
between: 1) FWHM H$\beta$ (the virial estimator of choice), 
2) $R_\mathrm{Fe{\sc ii}} = I($Fe{\sc ii}$\lambda$4570/I(H$\beta$),
3) soft X-ray photon index $\Gamma_\mathrm{soft}$ see also  \citep{wangetal96}
and 4) velocity shift of the C{\sc iv}$\lambda$1549  centroid at FWHM. Results so far are 
consistent with the assumption that source occupation
in 4DE1 has a physical basis with source Eddington ratio as the principal
driver \citep{marzianietal01}.

\begin{figure}
\begin{center}
\includegraphics*[width=13.25cm,angle=0]{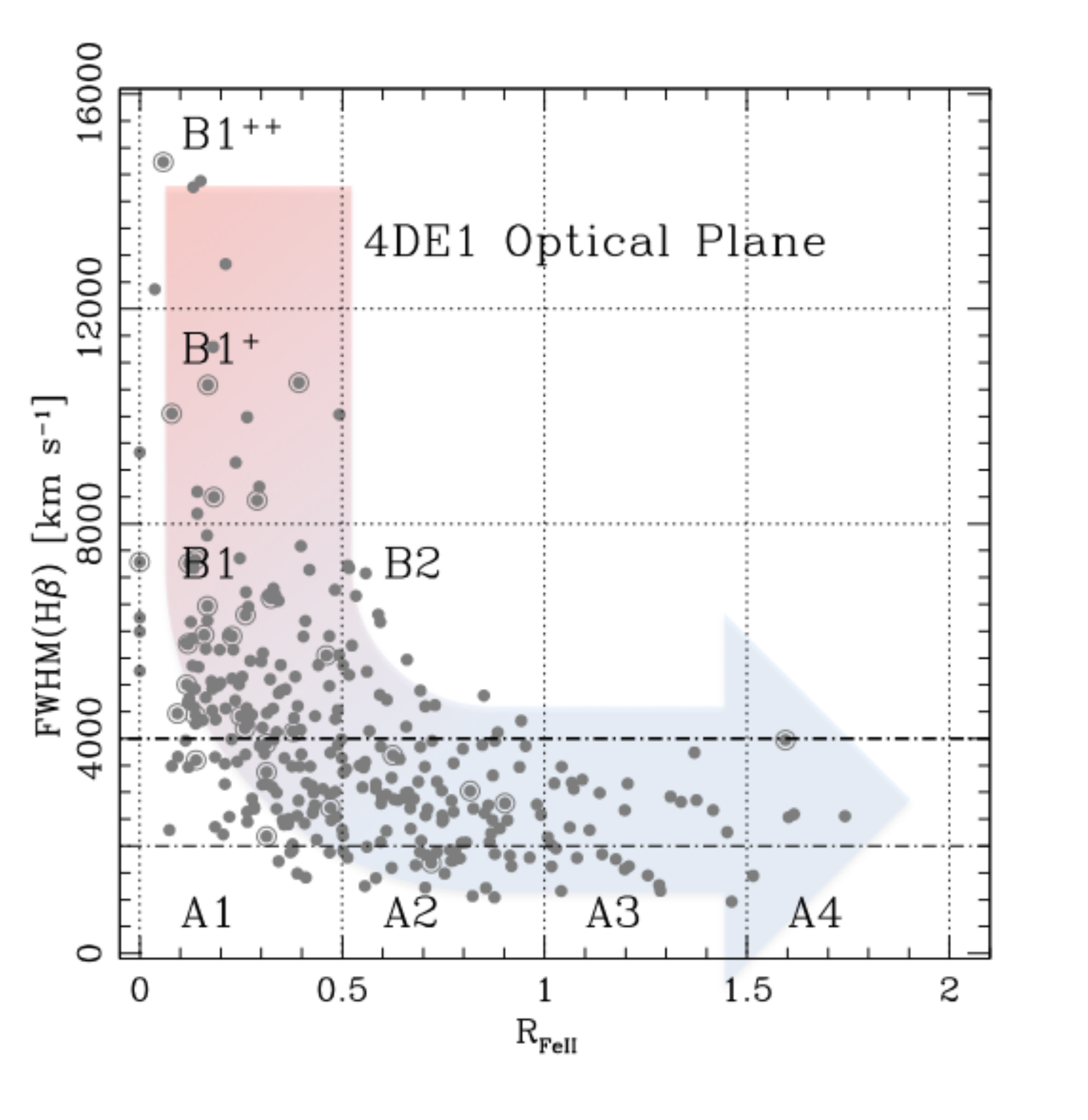}
\end{center}
\caption{The optical plane of the 4D eigenvector 1 parameter space, using data from \cite{zamfiretal10}. The plane has been subdivided in two regions occupied by Pop. A and B. A finer subdivision in spectral types (bins in FWHM and $R_\mathrm{Fe{\sc ii}}$) is also shown. The halftone arrow symbolizes several trends along the  sequence of sources in the plane: among them, the increase in high ionization line blueshift and in Eddington ratio toward extreme Pop. A sources (A3 and A4). Circled sources are radio-loud according to the criteria defined by  \citet{zamfiretal08}.    \label{fig:e1}}
\end{figure}

How can a context involving diagnostic measures help us derive more
accurate black holes masses? Can it
also help us to compare the relative merits of H$\beta$\ and Mg{\sc ii}$\lambda$2800 as virial 
estimators? Figure \ref{fig:e1} shows the optical
plane of 4DE1 which plots source FWHM H$\beta$ vs.  $R_\mathrm{Fe{\sc ii}} $\ measures. 
It is shown here for a large magnitude
limited sample of bright SDSS quasars \citep{zamfiretal10} where a clear
sequence of source occupation is observed. Sources with FWHM H$\beta <$4000 
km s$^{-1}$\ (population A) usually show a symmetric unshifted H$\beta$\ profile 
while sources with FWHM H$\beta >$4000 km s$^{-1}$\ (population B) require a 
double Gaussian model to describe the line. Pop A sources show Lorentz-like 
profiles so FWHM H$\beta$\ derived from Gaussian fits
to the line will result in overestimation of $M_\mathrm{BH}$. Pop B sources fit with a
single function and FWHM (or $\sigma$) measured from the fit, will result in serious
overestimation of $M_\mathrm{BH}$. Pop. B includes
the majority of radio-loud quasars so this overestimation often affects comparisons of
$M_\mathrm{BH}$ for radio-quiet and radio-loud
quasars. The double Gaussian required  to fit Pop. B sources involve a broad relatively
unshifted BLR component (BC) plus a
very broad (FWHM$\sim$10000km s$^{-1}$) and redshifted ($\Delta v_\mathrm{r}$=1000 -- 2000 km s$^{-1}$) VBC component. We have no choice but to adopt the unshifted component 
as the virial estimator -- but FWHM of this component is often  broadened by the VBC if the profile is not modelled.  

Most spectra lack high enough S/N  to allow modelling or even recognition
of the composite profiles for the almost half of quasars that show pop B characteristics. 
This is where the 4DE1 formalism can help. We know where the sources with simple and complex spectra are located in 4DE1 space. We can bin the optical plane of
4DE1 as shown in Figure \ref{fig:e1}  allowing us to generate median composite spectra 
for quasars in each bin. We thus avoid binning together dissimilar spectra which 
might be reflecting different BLR physics. The
resultant composite spectra show much higher S/N facilitating modeling of 
line profiles and  more accurate measures of black hole mass and Eddington ratio.
We can map median $M_\mathrm{BH}$ and $L/L_\mathrm{Edd}$ for sources across the 
4DE1 optical plane. Binned median spectra also open
the door to a more refined comparison between H$\beta$ and Mg{\sc ii}$\lambda$2800 as
virial estimators.

\section{Comparing H$\beta$ and Mg{\sc ii}$\lambda$2800 as Virial Estimators}

H$\beta$ is well established as the virial estimator of choice at $z<$ 0.7 -- 0.9.
There is no suitable low redshift
alternative except for H$\alpha$ which can only be used for sources below
$z \approx 0.2$. Of course both can be followed
into the IR, H$\beta$\ can  be used up to z=3.7. Suitably high S/N single spectra
and luminosity binned composite spectra suggest that the effect of nonvirial components
becomes more serious at high redshift \citep{sulenticetal06}. $M_\mathrm{BH}$ has
likely been overestimated for a large
fraction of high $z$ quasars. Such IR spectra are sufficiently costly in
telescope time that they are not likely
to provide measures for more than modest handfuls of 
sources \citep{marzianietal09}. Mg{\sc ii}$\lambda$2800 is a low
ionization resonance doublet that might reasonably be expected to show
similar propertied to H$\beta$ \citep{grandiphillips79,netzer80} but
with the added difficulty of narrow and broad-line absorption in an
unknown fraction of sources -- possibly more
common in Pop. A quasars.

SDSS has opened the door to detailed statistical studies of Mg{\sc ii}$\lambda$2800 and to comparison with H$\beta$\ \citep[e.g.,][]{trakhtenbrotnetzer12}. 
We find 680 sources in SDSS DR8 with spectra that provided high resolution
line profiles for both H$\beta$
and Mg{\sc ii}$\lambda$2800. They are observed in the redshift range from $z$ = 0.4 -- 0.75. 
There are more sources in this redshift range
but their spectra are too noisy to permit reliable 4DE1 spectral bin
assignments. Bright sources in this redshift
range cover a limited range of source luminosity. This means that median
composite spectra that we compute
for this bright quasar sample will reflect properties of sources within 1
dex of  $\log L_\mathrm {BOL} \approx$ 46. This is not bad
because it represents typical intermediate luminosity sources. The
composites trace a supposedly stable profile that is shared by most sources. They are conceptually different from  rms spectra computed for single sources in reverberation studies. 
Our assumption is that a large numbers of sources scatter around a well defined median/average in each 4DE1 bin. Using this procedure we are able to generate composite spectra for the 8 most
occupied bins in the 4DE1 optical plane.

\begin{figure}
\begin{center}
\includegraphics*[width=13.25cm,angle=0]{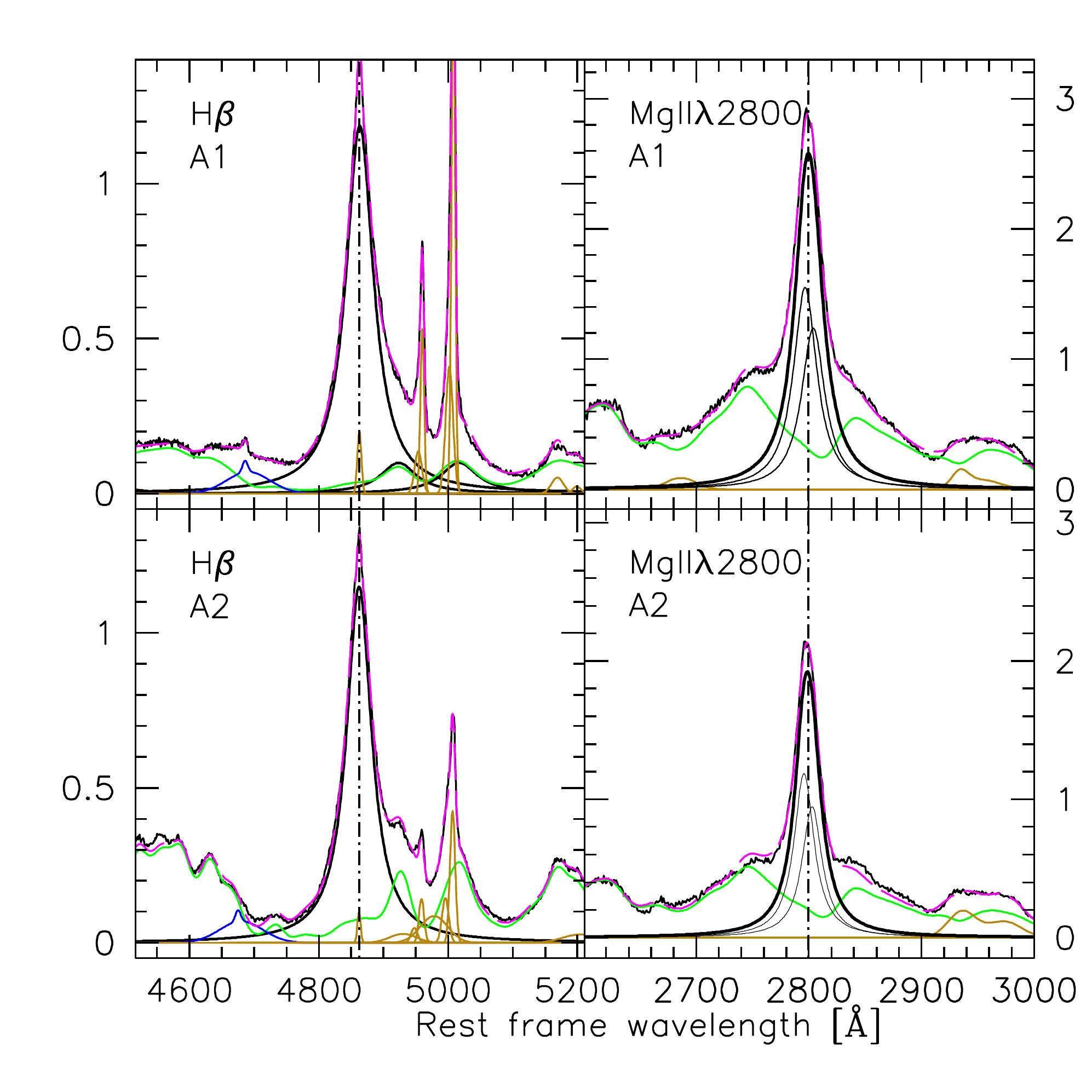}
\end{center}
\caption{Continuum subtracted spectral regions of H$\beta$ (left) and of Mg{\sc ii}$\lambda$2800\ (right) for spectral types A1 (top) and A2 (bottom). Abscissa is rest frame wavelength in \AA\ and ordinate is continuum-normalized intensity. Orange lines trace narrow lines and Fe{\sc i} emission  at $\approx$ 2900 \AA. The green line represents Fe{\sc ii} emission. Thick solid lines show  the broad components of H$\beta$ (left) and of Mg{\sc ii}$\lambda$2800.    \label{fig:a1a2}}
\end{figure}

\section{Comparison for Spectral Bins A1 and A2}

Figure \ref{fig:a1a2} shows median composite spectra for H$\beta$\ and Mg{\sc ii}$\lambda$2800 in 
bins A1 and A2. A2 is the most populated bin in Population A (48\% of the sample) with 
50\% of the Pop. A sample. The A bins are largely a sequence of increasing
$R_\mathrm{Fe{\sc ii}}$\ that subsumes the region formally defined for NLSy1 sources. 
Given the 1 \AA\ resolution of SDSS spectra we are forced to 
model Mg{\sc ii}$\lambda$2800 as a 2793/2806$\AA$ doublet of fixed ratio  1.25.
The composites in Figure 2 show {\sc iraf specfit} 
results superimposed including: 1) Fe{\sc ii} modelling, 2) fits to narrow
[O{\sc iii}]4959,5007 and H$\beta$, 3) He4686
and 4) broad H$\beta$. When we use {\sc iraf specfit} we must specify what the
program should  look for or it will never find a reasonable solution with so many free parameter lines and components present. In  this case we assume that both
components of Mg{\sc ii}$\lambda$2800 are unshifted Lorentz profiles. This provides a good fit judging from the residuals shown at the
bottom of the fits in Fig. 2. Gaussian models provide an inferior fit
consistent with what we always find for H$\beta$.
The main result of this comparison indicates that FWHM Mg{\sc ii}$\lambda$2800 is 20\%\ smaller
than FWHM H$\beta$\ meaning that Mg{\sc ii}$\lambda$2800 used
as the virial estimator will yield systematically smaller black holes
masses than H$\beta$\ \citep[in agreement with the BLR self-shielding model of][]{gaskelletal07}. In general if one employs
Mg{\sc ii}$\lambda$2800 one must pay attention to the resolution of the spectra being
measured relative to the separation of the Mg{\sc ii}$\lambda$2800 doublet.
Otherwise one will overestimate $M_\mathrm{BH}$ and one will find less than a 20\%
FWHM difference relative to H$\beta$\ (e.g. in bin A1
FWHM H$\beta$=3180 while for Mg{\sc ii}$\lambda$2800 the full profile FWHM $\approx$ 3040 km s$^{-1}$ while individual multiplets show FWHM Mg{\sc ii}$\lambda$2800 $\approx$ 2710 km s$^{-1}$).
Note that in this comparison  a black hole mass estimation also requires a
value of $r$\ which is obtained from extrapolation
of the Kaspi relation. The luminosities used to derive $r$ will not be the
same for the two lines. $L$\ is derived from the
continuum near each line: 3100 \AA\ and 5100 \AA\ for Mg{\sc ii}$\lambda$2800 and H$\beta$
respectively. Best estimate median log $M_\mathrm{BH}$ values
for bins A1/A2 are 8.67/8.57 and 8.62/8.45 for H$\beta$\ and Mg{\sc ii}$\lambda$2800
respectively (solar units). Best log Eddington ratio estimates
for A1/A2 are -0.60/-0.53  and -0.56/-0.40 for H$\beta$\ and Mg{\sc ii}$\lambda$2800
respectively.   It is assumed that the smaller
of the two FWHM values (e.g. FWHM Mg{\sc ii}$\lambda$2800 BC which is a single VBC corrected term of the doublet) is the best
estimate with  0.5 dex uncertainty.

\begin{figure}
\begin{center}
\includegraphics*[width=13.25cm,angle=0]{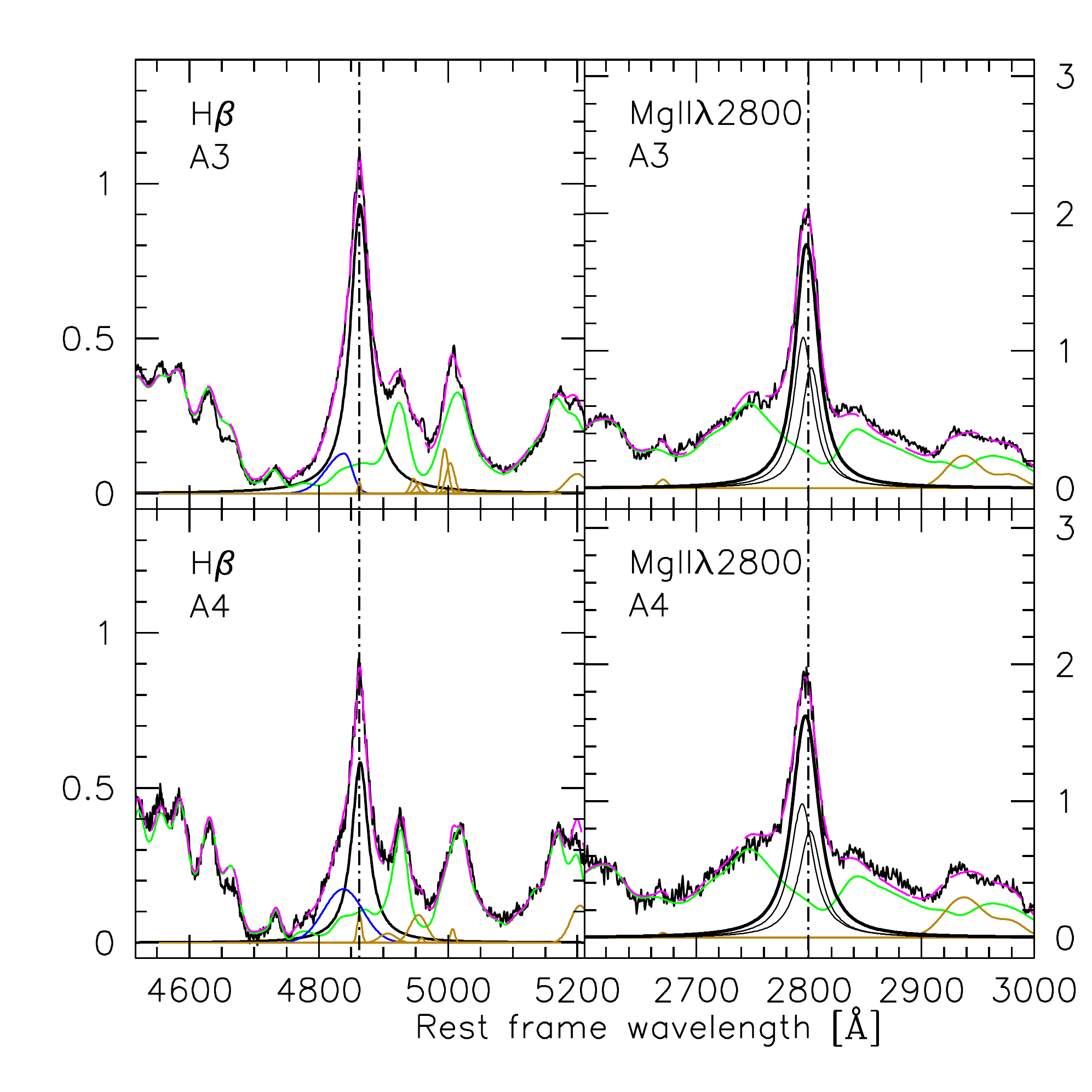}
\end{center}
\caption{Continuum subtracted spectral regions of H$\beta$ (left) and of Mg{\sc ii}$\lambda$2800\ (right) for spectral types A3 (top) and A4 (bottom). Abscissa is rest frame wavelength in \AA\ and ordinate is continuum-normalized intensity. Orange lines trace narrow line and FeI emission, the latter at $\approx$ 2900 \AA. The green line represents Fe{\sc ii} emission. Thick solid lines show  the broad components of H$\beta$ (left) and of Mg{\sc ii}$\lambda$2800.    \label{fig:a3a4}}
\end{figure}

\section{Comparison for spectral bins A3 and A4}

Figure \ref{fig:a3a4} shows  composite spectra for bins A3 and A4 which involve 43 and
15 sources respectively. Results of best {\sc specfit}
models are superimposed with residuals displayed below. In many ways these
bins represent the most extreme quasars in a low  redshift sample. They show the strongest Fe{\sc ii} emission ($R_\mathrm{Fe{\sc ii}} >$1.0), a
C{\sc iv}$\lambda$1549 blueshift/asymmetry with some shift amplitudes
exceeding 1000km s$^{-1}$  and a soft  X-ray excess ($\Gamma_\mathrm{soft}>$ 2.7). The  
well known NLSy1 quasar I Zw 1 is found in bin A3.
Computed $ \log M_\mathrm{BH}$ values for these sources A3/A4 (8.33/8.27 and 8.41/8.56
for H$\beta$ and Mg{\sc ii}$\lambda$2800 respectively) coupled with
source luminosities lead us to infer that these sources are radiating
closest to the Eddington limit (A3/A4 --0.28/--0.18
and --0.37/--0.47 for H$\beta$ and Mg{\sc ii}$\lambda$2800 respectively).  See
\citet{marzianietal03d,marzianietal06} and communication in this proceeding for the cosmological
potential of such sources. These sources show surprisingly large scatter
in a FWHM Mg{\sc ii}$\lambda$2800 vs FWHM H$\beta$ plot and gave the
impression that the difference between FWHM H$\beta$ and FWHM Mg{\sc ii}$\lambda$2800 was
converging towards zero  at low FWHM
values (see e.g. \citealt{wangetal09}). In fact they show anomalous
properties and are sources where Mg{\sc ii}$\lambda$2800 cannot be trusted
as a virial estimator. We find that indeed FWHM Mg{\sc ii}$\lambda$2800 is equal 
too or even slightly greater than FWHM H$\beta$\ -- the only bins
where this is observed. In additional the median Mg{\sc ii}$\lambda$2800 profiles for these
two bins show a systematic blueshift that is
largest in bin A4 (--265km s$^{-1}$). The blueshift is attributed to a wind or
outflow associated with the high $L/L_\mathrm{Edd}$
values for these sources. The profile of  H$\beta$ for these two bins also shows a
small blueshifted component perhaps related to the
same outflow process. Since it can be well modeled and does not affect
H$\beta$ at the FWHM level we conclude that H$\beta$\
can be trusted as a virial estimator.

\section{Comparison for spectral B bins}

Figure 4 shows composite spectra for bins B1 and B1+ with associated
specfits and residuals superposed and
displayed below, respectively. Population B represents a series of bins
(52\% of our sample) with increasing
FWHM H$\beta$. They share a constant range of $R_{FE}<$0.5  except for bin B2
which shows a mean $R_\mathrm{Fe{\sc ii}}  \approx 0.5 - 1.0$. Only results for
Bins B1 and B1+ are shown. With 218 and 115 sources, respectively, they
represent 90\% of the population B sample.
These sources require a double Gaussian (BC+VBC) fit to both H$\beta$\ and
Mg{\sc ii}$\lambda$2800. Given the large uncertainies about: 1) the widths
of both components, 2) the BC/VBC intensity ratio and  3) the amplitude of
the VBC redshift, these fits are the least well constrained. We know that the BC/VBC intensity ratio can show a wide range \citep{marzianietal09}. Some sources in fact
show a ratio near to zero (e.g. PG1416-129, \citealt{sulenticetal00b} and
3C110 \citealt{marzianietal10}). While compiling our SDSS
sample for the H$\beta$\ -- Mg{\sc ii}$\lambda$2800 comparison we found additional examples of
what we call super VBC quasars. Figure 5 shows the H$\beta$
and Mg{\sc ii}$\lambda$2800 profiles for PG1201+436 which is clearly a case where the BC/VBC
ratios for H$\beta$\ and Mg{\sc ii}$\lambda$2800 profiles are very different.
Bins B1 and B1+ show $M_\mathrm{BH}$ values B1/B1+ of 9.15/9.29 and 8.98/9.06 
for H$\beta$\ and Mg{\sc ii}$\lambda$2800 respectively. The estimated Eddington ratios are
-1.13/-1.44  and -0.96/-1.21 for H$\beta$\ and Mg{\sc ii}$\lambda$2800 respectively. Since the
Mg{\sc ii}$\lambda$2800 VBC is weaker in the Mg{\sc ii}$\lambda$2800 composites we consider the
$M_\mathrm{BH}$ and $L/L_\mathrm{Edd}$ values for that line to be more reliable.

\begin{figure}
\begin{center}
\includegraphics*[width=13.25cm,angle=0]{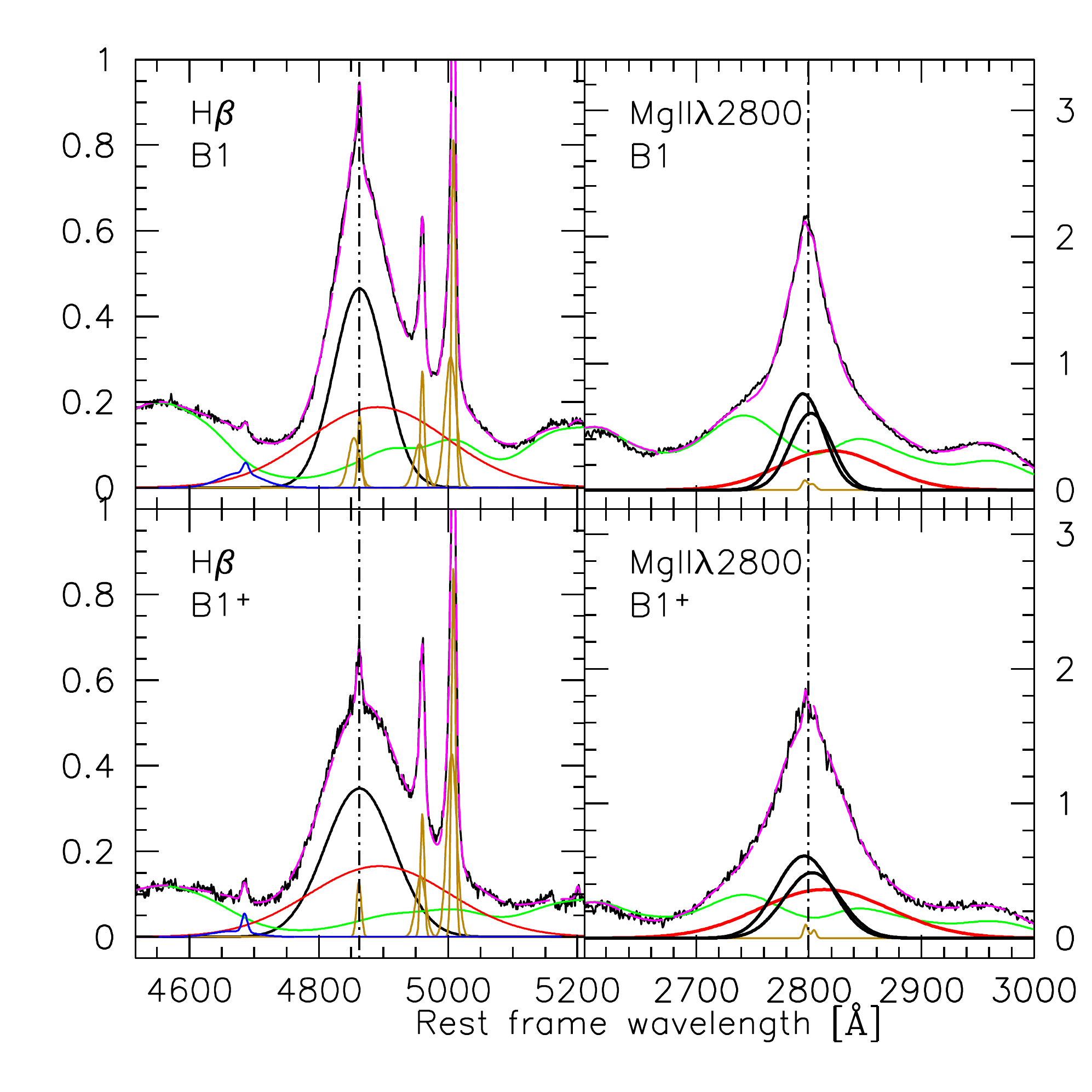}
\end{center}
\caption{Continuum subtracted spectral regions of H$\beta$ (left) and of Mg{\sc ii}$\lambda$2800\ (right) for spectral types B1 (top) and B1$^+$ (bottom). Abscissa is rest frame wavelength in \AA\ and ordinate is continuum-normalized intensity. Orange lines trace narrow line emission, the latter at $\approx$ 2900 \AA. The green line represents Fe{\sc ii} emission. Thick solid lines show  the broad components of H$\beta$ (left) and of Mg{\sc ii}$\lambda$2800.    \label{fig:b1b1p}}
\end{figure}

\begin{figure}
\begin{center}
\includegraphics*[width=13.25cm,angle=0]{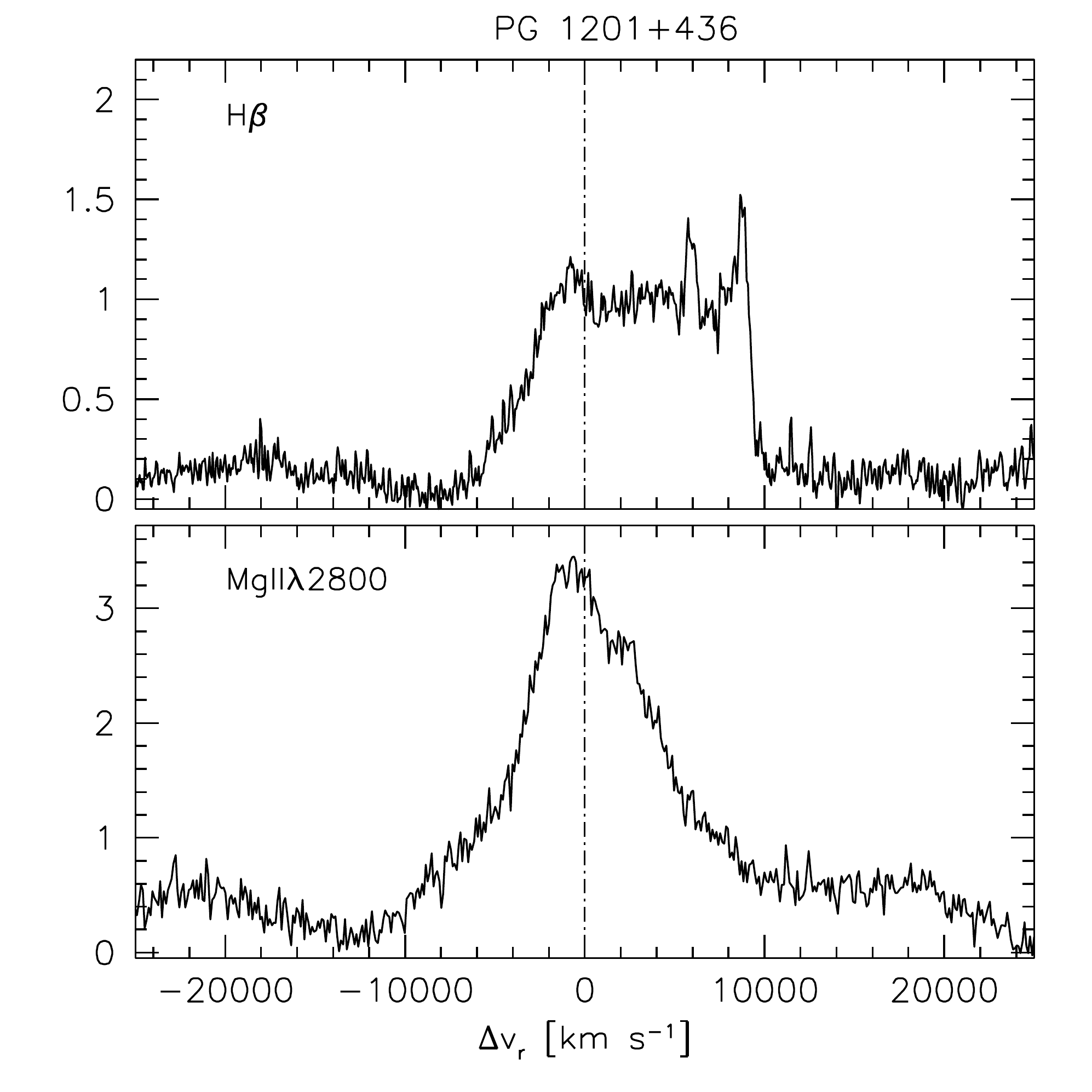}
\end{center}
\caption{Continuum subtracted spectral regions of H$\beta$ (top) and of Mg{\sc ii}$\lambda$2800\ (bottom) for source PG 1201+436. Abscissa is radial velocity difference in km s$^{-1}$\ with respect to line rest frame wavelength and ordinate is continuum-normalized intensity.   \label{fig:pg1201}}
\end{figure}

\section{Conclusions}

We are still struggling with $M_\mathrm{BH}$ (and consequently $L/L_\mathrm{Edd}$) estimates
largely with $\pm$1dex uncertainties. However we are laying the foundation
for estimations with uncertainties of a few 0.1dex--especially bin-to-bin
uncertainties which are almost as valuable as absolute ones. The path
towards more accurate estimates lies within a context like 4DE1 where
spectral binning can greatly increase S/N of the line measures. The latest
bin results (for log $L_\mathrm{bol} = 46 \pm $0.5  --a typical quasar luminosity) ) show a 
trend in the 4DE1 optical plane of decreasing
$M_\mathrm{BH}$ from bins B1+/B1++ (9.1) to  bin A4 (8.3). Resultant $L/L_\mathrm{Edd}$ values
increase  from --1.5 to --0.2.  The full $M_\mathrm{BH}$\ range is likely 7.0 -- 9.5
(with $L/L_\mathrm{Edd}$\ range from --2.0  to  0.0 in logarithm) but smaller bins and a larger range
in source luminosity is needed to explore it. If $M_\mathrm{BH}$\ grows with time via
merging and accretion then the 4DE1 trend may also represent an
evolutionary sequence with quasars in this sample middle-aged and the
youngest sources in bins A3/A4 radiating at or near the Eddington limit.
%%%%%%%%%%%%%%%%%%%%%%%%%%%%%%%%%%%%%%%%%%%%%%%%%%%%%%%%%%%%%%%%%%%%%%%%%%%%%
%% Appendices
% \appendix

\paragraph{Acknowledgements} {\small Funding for the SDSS and SDSS-II has been provided by the Alfred P. Sloan Foundation, the Participating Institutions, the National Science Foundation, the U.S. Department of Energy, the National Aeronautics and Space Administration, the Japanese Monbukagakusho, the Max Planck Society, and the Higher Education Funding Council for England. The SDSS Web Site is http://www.sdss.org/.

\small The SDSS is managed by the Astrophysical Research Consortium for the Participating Institutions. The Participating Institutions are listed on the webpage http://www.sdss.org/collaboration/credits.html.}

\bibliographystyle{model1b-num-names} 
%\bibliography{biblioletter}

%\clearpage

\end{document}